\documentclass[10pt,conference,comsocconf]{IEEEtran}

\usepackage{blindtext}
\usepackage{graphicx}

\usepackage{multirow} 
\usepackage{tikz} 
\usetikzlibrary{calc,fit,shapes}
\usetikzlibrary{hobby,backgrounds,trees}
\usepgflibrary{arrows}
\usepackage{varwidth}
\usetikzlibrary{positioning,arrows}

\usepackage{xargs}                      
\usepackage[colorinlistoftodos,prependcaption,textsize=small]{todonotes}
\newcommandx{\pablo}[2][1=]{\todo[inline,linecolor=green,backgroundcolor=green!25,bordercolor=green,#1]{#2}}
\newcommandx{\ordonez}[2][1=]{\todo[inline,linecolor=red,backgroundcolor=red!25,bordercolor=red,caption={},#1]{#2}}
\newcommandx{\oadamuz}[2][1=]{\todo[inline,linecolor=yellow,backgroundcolor=yellow!25,bordercolor=yellow,caption={},#1]{#2}}

\ifCLASSINFOpdf
\else
\fi
\usepackage{fancyhdr}

\begin{document}

\pagestyle{fancy}
\fancyhf{} 
\renewcommand{\headrulewidth}{0pt}
\fancyhead{} 
\fancyhead[C]{\fontsize{8}{10} \selectfont This article has been accepted for publication in the European Conference on Networks and Communications (EuCNC), 2018. } 

%
\title{The Creation Phase in Network Slicing: From a Service Order to an Operative Network Slice}
%
%
%
\author{\IEEEauthorblockN{Jose Ordonez-Lucena\IEEEauthorrefmark{1}\IEEEauthorrefmark{2}, Oscar Adamuz-Hinojosa\IEEEauthorrefmark{1}\IEEEauthorrefmark{2}, Pablo Ameigeiras\IEEEauthorrefmark{1}\IEEEauthorrefmark{2}, Pablo Mu\~noz\IEEEauthorrefmark{1}\IEEEauthorrefmark{2}, Juan J. Ramos-Mu\~noz\IEEEauthorrefmark{1}\IEEEauthorrefmark{2}, \\ Jes\'us Folgueira Chavarria\IEEEauthorrefmark{3}, Diego Lopez\IEEEauthorrefmark{3}}

\IEEEauthorblockA{\IEEEauthorrefmark{1}Research Center on Information and Communication Technologies, University of Granada.}
\IEEEauthorblockA{\IEEEauthorrefmark{2}Department of Signal Theory, Telematics and Communications, University of Granada.}
\IEEEauthorblockA{\IEEEauthorrefmark{3} Telefonica I+D-Global CTO.\\ Email: \{jordonez, oadamuz, pameigeiras, pabloml, jjramos\}@ugr.es\IEEEauthorrefmark{1}\IEEEauthorrefmark{2}  \{jesus.folgueira, diego.r.lopez\}@telefonica.com\IEEEauthorrefmark{3}}
}

%
%

\markboth{Journal of \LaTeX\ Class Files,~Vol.~6, No.~1, January~2007}%
{Shell \MakeLowercase{\textit{et al.}}: Bare Demo of IEEEtran.cls for Journals}
%



\IEEEoverridecommandlockouts

\IEEEoverridecommandlockouts

\IEEEpubid{\begin{minipage}{\textwidth}\ \\[10pt]
\centering\normalsize{978-1-5386-1478-5/18/\$31.00~\copyright~2018 IEEE}
\end{minipage}}

\makeatletter
\def\ps@IEEEtitlepagestyle{
  \def\@oddfoot{\mycopyrightnotice}
  \def\@evenfoot{}
}
\def\mycopyrightnotice{
  {\footnotesize
  \begin{minipage}{\textwidth}
  \centering
 ~\copyright~2018 IEEE. Personal use of this material is permitted. Permission from IEEE must be  obtained for all other uses, in any current or future media, \\ including reprinting/republishing this material for advertising or promotional purposes, creating new  collective works, for resale \\ or redistribution to servers or lists, or reuse of any copyrighted component of this work in other works.
\end{minipage}
  }
}

\maketitle

\begin{abstract}
Network slicing is considered a key mechanism to serve the multitude of tenants (e.g. vertical industries) targeted by forthcoming 5G systems in a flexible and cost-efficient manner. In this paper, we present a SDN/NFV architecture with multi-tenancy support. This architecture enables a network slice provider to deploy network slice instances for multiple tenants on-the-fly, and simultaneously provision them with isolation guarantees. Following the Network Slice as-a-Service delivery model, a tenant may access a Service Catalog, selecting the slice that best fits its needs and ordering its deployment. This work provides a detailed view on the stages that a network slice provider must follow to deploy the ordered network slice instance, accommodating it into a multi-domain infrastructure, and putting it operative for tenant's consumption. These stages address critical issues identified in the literature, including (i) the mapping from high-level service requirements to network functions and infrastructure requirements, (ii) the admission control, and (iii) the specific information a network slice descriptor should have. With the proposed architecture and the recommended set of stages, network slice providers can deploy (and later operate) slice instances with great agility, flexibility, and full automation.

\end{abstract}


%
\IEEEpeerreviewmaketitle

\section{Introduction}

The ongoing digital transformation is geared towards the integration of vertical industries into an ecosystem boosting technical and business innovation. This may bring a multitude of new vertical-driven use cases and application scenarios, with very distinct requirements. Current research efforts focus on finding ways to accommodate them on the same infrastructure in a flexible, agile, and cost-efficient manner. Network slicing will be key for this end. Leveraging network softwarization technologies such as Software Defined Networking (SDN) and Network Functions Virtualization (NFV), network slicing aims to logically split an infrastructure into a set of self-contained programmable network instances, each customized to only serve the particular needs of a given use case. The shared and multi-domain nature of the infrastructure on top of which these Network Slice (NSL) instances run makes isolation a capital requirement for network slicing.

Network slicing has brought the attention of the research community. Many standardization bodies and Fora have addressed this concept, including NGMN, IETF, ONF, and 3GPP. In \cite{nfv:eve012}, ETSI NFV provides an insight into the different views that some of these organizations have about slicing, analyzing how their visions match with the NFV constructs. 

Network slicing is claimed to unlock new business opportunities, with flexible service delivery models. One of them is Network Slicing as-a-Service \cite{Zhou:NetworkSlicing}. This service delivery model enables an NSL provider (e.g. network operator) to deploy customized NSL instances for their clients (e.g. verticals) on request, and deliver them as a service. These clients, taking the role of NSL tenants, may in turn use the purchased NSL instances to deploy their business services for their own clients.  This empowers recursive business models (e.g. Business-to-Business-to-X models), with multiple actors providing services at different positions in the value chain. 

In our previous work \cite{ordonez-NSL-architecture}, we proposed an SDN/NFV-based architecture enabling operation of NSL instances with recursiveness, multi-tenancy and multi-domain support. Although these issues have been addressed in architectural solutions proposed in different 5G-PPP projects (e.g. 5G-Crosshaul, 5GNORMA, 5GEx, etc.), none of them consider the isolation as the first criteria for architecture design. This have lead to solutions that do not address all the isolation properties necessary in slicing: \textit{performance, security, privacy, and management isolation}. Unlike those proposals, our solution satisfies each of these isolation properties while being compliant with ETSI NFV information model. For this end, two architectural enhancements are considered with respect to the NFV framework \cite{nfv:man001}: the decomposition of the NFV Orchestrator (NFVO) into resource and network service orchestration blocks, and the inclusion of a Tenant SDN Controller. The results derived from this work have contributed in ongoing standardization efforts, including those conducted by ETSI NFV \cite{nfv:eve012} and IETF \cite{ietf}. 

\IEEEpubidadjcol

The vision given in \cite{ordonez-NSL-architecture} focused on the \textit{run-time phase}, considering the NSL instances were operative and leased out to their tenants. However, the \textit{creation phase} was omitted. In this phase, a tenant requests a NSL from a catalog, and orders its instantiation. The creation phase brings new challenges, including the translation of tenant-specific service requirements into network functions and infrastructure requirements, the specification of an NSL descriptor, and the admission control. These and other aspects have been identified in \cite{Foukas:NetworkSlicing} as still open issues in the context of network slicing. Addressing them is thus essential to make a complete network slicing solution. 

In this paper, we concentrate on the creation phase of network slicing, complementing the run-time phase addressed in our previous work. The main objective is to provide an insight into the procedures and mechanisms required to make the deployment of NSLs more flexible, agile, and automated from the perspective of both the NSL provider and the tenant. To incorporate these mechanisms and procedures, we extend our SDN/NFV-based architecture with two new functional blocks: the NSL Manager and the NSL Orchestrator. In the context  of this architecture, we identify the stages the NSL provider shall follow for completing a catalog-driven NSL deployment. In each stage, we specify the input/output information, the steps involved, and the role that each functional block plays.

The remainder of this article is as follows. Section \ref{sec:NFV} shows how the concept of ETSI NFV network service is key to provide a resource-centric view of an NSL. Section \ref{sec:run-time} describes the slicing architecture, with focus on the new functional blocks. Section \ref{sec:deployment-time} provides a detailed view on the creation phase, on a step-by-step basis. Finally, Section \ref{sec:conclusions} summarizes the main conclusions of this work.\\

\section{NFV Network Services and Network Slices}\label{sec:NFV}
The concept of Network Service (NS) introduced by ETSI NFV is key for network slicing. NSLs leverage the capabilities offered by NSs to satisfy the network requirements of the use cases they accommodate. From a resource-centric viewpoint, an NSL instance may be composed of one or more NS instances. Particularly, three scenarios can be considered: 

\begin{itemize}
\item [(a)] The NSL instance consists of an instance of a simple NS.
\item [(b)] The NSL instance consists of an instance of a composite NS.
\item [(c)] The NSL instance consists of a concatenation of simple and/or composite NS instances.
\end{itemize}

\begin{figure}[htbp!]
\centering 
\includegraphics[width=0.29\textwidth]{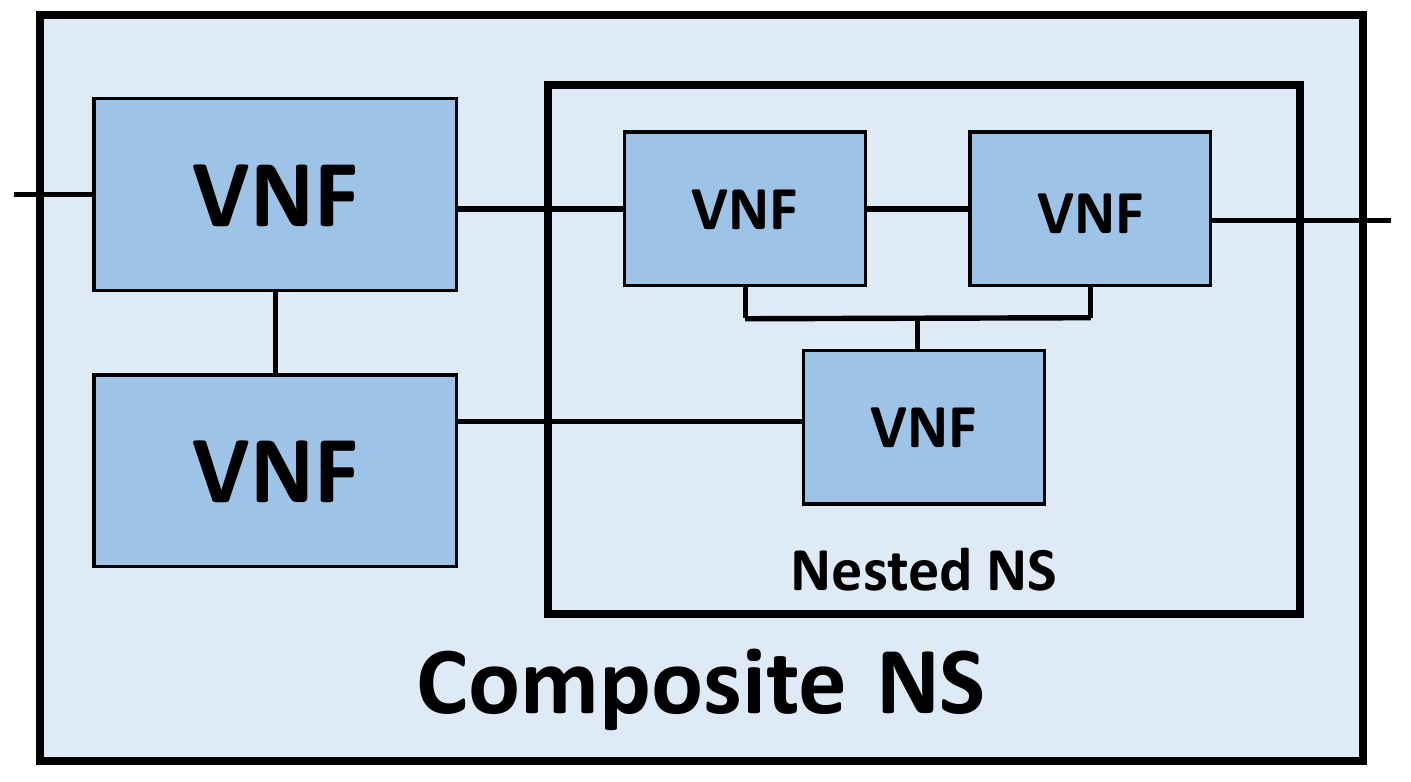}
\caption{An example of a composite NS. This NS consists of two VNFs and one simple NS.  
}
\label{fig:NSLComposition}
\end{figure} 

A simple NS includes one or more Virtualized Network Functions (VNFs), and virtual links providing connectivity between them.  In search of modularity and recursiveness, the NFV framework provides the ability to include in the design of an NS one or more nested NSs. The result is a composite NS (see Fig. \ref{fig:NSLComposition}).

According to ETSI NFV, an NS instance is deployed from an NS descriptor. An NS descriptor is a deployment template used for creating and operating instances of an NS. The NS descriptor provides a list of pointers to the VNF descriptors of the constituent VNFs, and additional information on connectivity between them. In case of a composite NS, the corresponding NS descriptor also references the NS descriptor(s) of the nested NS(s).

A key mechanism in the NS descriptor is NS flavoring. NS flavoring enables customizing the deployment of an NS instance, in terms of \textit{functionality} and \textit{performance}. As stated in \cite{nfv:ifa014}, an NS descriptor consists of one or more NS flavors, each specifying a different deployment configuration for the NS. Selecting an NS flavor within the NS descriptor enables selecting the VNFs and virtual links to be deployed as part of the NS, and hence the features to be activated for that NS. 

A given NS flavor includes one or more NS Instantiation Levels (NS-ILs), each specifying a possible option of instantiating the NS using this flavor. An NS instance resulting from a NS-IL can only include instances of those VNFs and virtual links that have been declared in the flavor. The goal of a NS-IL is to describe how to deploy each constituent VNF and virtual link. To that end, an NSL-IL contains the following:

\begin{itemize}
\item For each VNF to be used for the NS instance, the NS-IL specifies the number of instances to be deployed, their resource levels (i.e. the level of resources to be allocated for each instance), and their applicable affinity/anti-affinity rules. Currently, the reliability requirements of a VNF (e.g. the subset of instances to serve as backup, if high availability hardware/software is required for any instance, etc.) are not part of the NS-IL, although their inclusion is expected for the NFV Release 3 \cite{nfv:eve012}.
\item For each virtual link to be used for the NS instance, the NS-IL specifies transport reliability and the bitrate requirements.
\end{itemize}

According to the mentioned ideas, a triplet (NS descriptor ID, NS Flavor ID, NS-IL ID) provides a complete resource-centric description of an NS instance. The second term indicates the subset of VNFs and virtual links to be deployed for the NS, and hence the functionality selected for the NS. The third term specifies how instantiating each of those VNFs and links, thus setting the level of performance of the NS.

As seen, NS flavoring is key for slicing, as it enables selecting only the needed capabilities within an NS for a given NSL. To provide a complete resource-centric description of an NSL instance, it is required to specify which triplet is used to instantiate each constituent NS. For this end, we introduce the concept of NSL Instantiation Level (NSL-IL). The NSL-IL is an information element that provides a (list of) pointer(s) to the triplet(s) of the constituent NS instance(s). This means that if an NSL instance have  $M$ NS instances - see (c) -, then the NSL-IL will refer to the $M$ triplets used for their instantiation.

\section{Network Slicing Architecture}\label{sec:run-time}

In this section, we describe a SDN/NFV based architecture for network slicing that extends our previous proposal \cite{ordonez-NSL-architecture}. Note that this architecture focuses on the transport and core network domains, omitting the RAN domain for simplicity. 

\begin{figure}[htpb!]
\centering 
\includegraphics[width=0.49\textwidth]{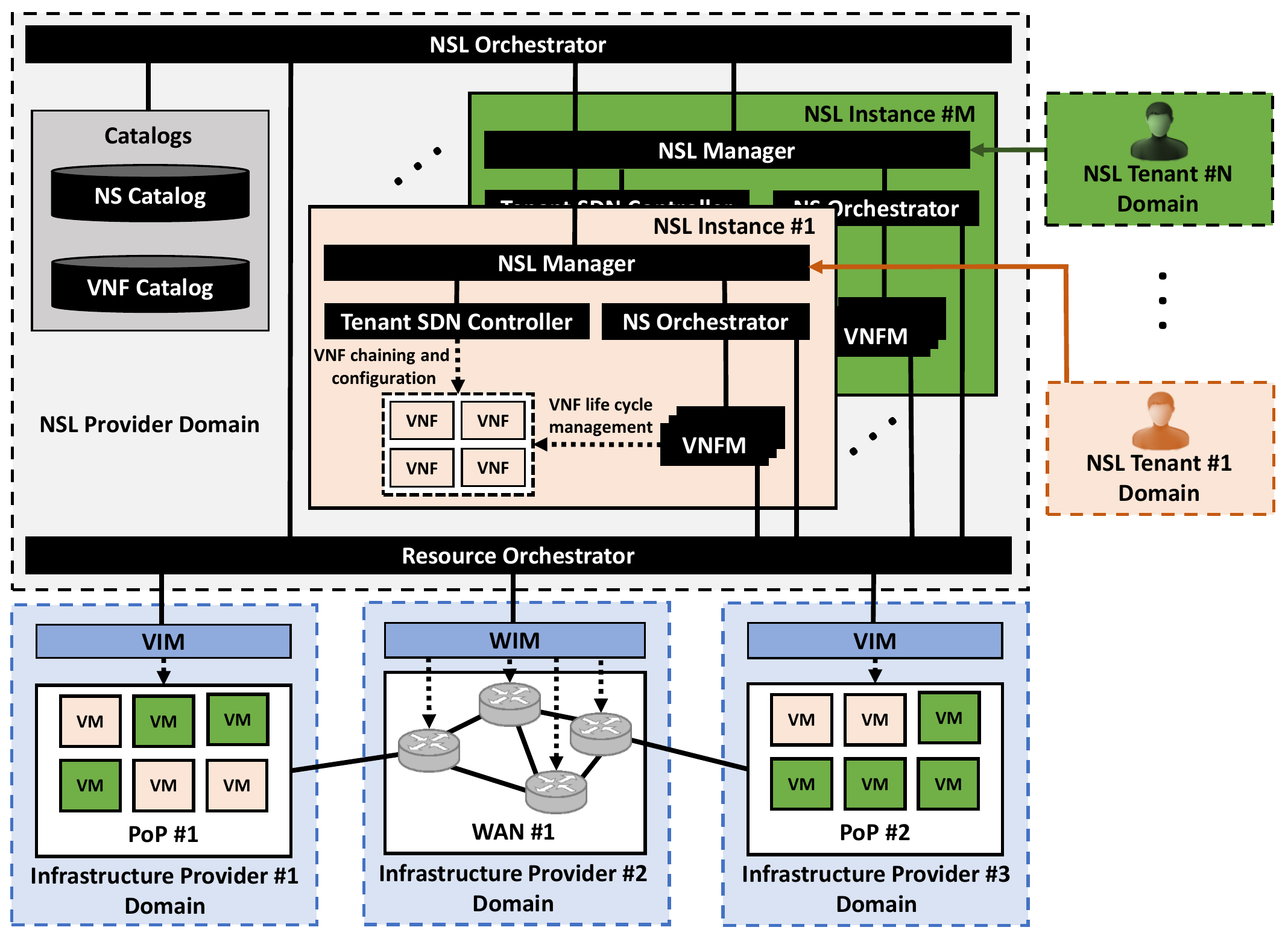}
\caption{SDN/NFV-based Network Slicing Architecture.}
\label{fig:architecture}
\end{figure}

As Fig. \ref{fig:architecture} shows, this architecture enables an NSL provider to simultaneously operate multiple NSL instances. These instances run on top of a common infrastructure that spans across multiple administrative domains, each belonging to a different infrastructure provider. This infrastructure, consisting of geographically distributed Points of Presence (PoPs) and Wide Area Networks (WANs) connecting them, enables multi-site deployments. To manage the resources of the PoP(s) and/or WAN(s) within its administrative domain, an infrastructure provider leverages the capabilities of a Virtual Infrastructure Manager (VIM) and/or WAN Infrastructure Manager (WIM), respectively.

The NSL provider, taking the role of an infrastructure tenant, rents the infrastructure resources owned by the underlying infrastructure providers, and uses them to provision the NSL instances. For this end, the NSL provider has a resource orchestration functional block. The Resource Orchestrator uses the finite set of resources that are at its disposal (the resources supplied by the underlying VIMs/WIMs), and dispatches them to the NSL instances in an optimal way. This optimization means that all the NSL instances are simultaneously provided with the resources needed to satisfy their (potentially diverging) requirements, while preserving their performance isolation. The resource requirements of  each NSL instance are stated by its NSL-IL (see Section \ref{sec:NFV}). 

An NSL instance uses its assigned resources to run instances of VNFs. These VNF instances are stitched together to build up the required NS instance(s), following the specificities given in the NSL-IL. At infrastructure level, note that VNF instances are executed on virtualization containers (e.g., virtual machines [VM], docker containers, unikernels, etc.). These virtualization containers are deployed inside one or more PoPs, according to the geolocation requirements of the VNFs. 

To preserve management isolation across NSL instances, each instance has its own management plane. This plane consists of four functional blocks: VNF Manager (VNFM), NS Orchestrator, Tenant SDN Controller, and NSL Manager.  

The VNFM(s) and the NS Orchestrator perform the required life cycle operations (e.g. instantiation, scaling, termination, etc.) over the instances of the VNFs and NS(s), respectively. Since these operations involve modifying the amount of resources to be allocated for those instances, an interplay between these functional blocks and the Resource Orchestrator is required. The Tenant SDN Controller performs VNF configuration and chaining in a programmatic manner. On one hand, this SDN Controller configures the VNF instances at application level, taking the role of an Element Manager (EM) \cite{nfv:man001}. On the other hand, it chains the VNF instances for NS construction, leveraging the forwarding capabilities provided by the data plane. Finally, the NSL Manager coordinates the operations and management data from both the Tenant SDN Controller and the NS Orchestrator, performing the fault, configuration, accounting, performance, and security management within the NSL instance. Additionally, it provides visibility and management capability exposure to external blocks. In this respect, note that the NSL Manager is of key importance for a NSL tenant. Each tenant consumes its NSL instance, and operates it at its convenience (within the limits agreed with the NSL provider) through the NSL Manager. By way of example, the tenant could use an SDN application in the NSL manager to programmatically modify the VNF chaining rules on-the-fly, according to its needs. 

Beyond the domain of an NSL instance, the NSL provider defines an NSL Orchestrator. This functional block plays a key role in the creation phase and the run-time phase. In the creation phase, it receives the order to deploy a NSL instance for a tenant, checks the feasibility of the order, and if feasible, triggers the instantiation of the NSL. For this end, it interacts with the Resource Orchestrator, and accesses the VNF and NS Catalogs. These catalogs contain VNF and NS descriptors, exposing the capabilities of all the VNFs and NSs that an NSL provider can select for the NSLs. At run-time, the NSL Orchestrator performs policy-based inter-slice operations. Particularly, it analyses the performance and fault management data received from the operative NSL instances to manage their Service Level Agreements. In case of Service Level Agreement violations, then the NSL Orchestrator decides which NSL instances need to be modified, and sends corrective management actions (e.g. scaling, healing, etc)  to their NSL Managers.

The interplay among the functional blocks described so far enables slicing. Abstraction is a key architectural principle for this end. Having different abstraction levels across functional blocks logically placed at different layers leads to a loosely coupled architecture. Each functional block is only responsible for a specific set of tasks, being they limited by the level of information the functional block understands. 

In our architecture, the VIM/WIM, the Resource Orchestrator, and the NSL/NS Orchestrator operate at different layers, and hence provide different abstraction levels. The Resource Orchestrator maintains a PoP resource map derived from the information provided by VIM(s) and WIM(s), including data on geolocation, capabilities\footnote{The capabilities of a PoP depend on the PoP setup (e.g. setup for high availability and fault resiliency, setup for high I/O processing, etc.)}, and resource state. The Resource Orchestrator abstracts this information to the NSL/NS Orchestrator, providing a resource-agnostic view of the set of reachable PoPs. This view only includes high-level information on the locations and capabilities of those PoPs, without any information on their resources, nor the VIM(s) responsible for their management.

\section{Network Slice Creation Phase}\label{sec:deployment-time} 

Section \ref{sec:run-time} focuses on the run-time phase of the network slicing concept, considering that the NSL instances are operative and being consumed by their tenants. However, prior to this phase, the creation phase occurs. This section concentrates on the creation phase, providing a detailed view on the  steps the NSL provider must follow to instantiate a NSL according to the specificities gathered in a catalog-driven service order. For better understandability, these steps have been grouped into five well-defined stages. These stages are described below.  

\subsection{Service ordering}\label{subsec:serviceordering}

\begin{figure}[htbp!]
\centering 
 \includegraphics[width=0.46\textwidth]{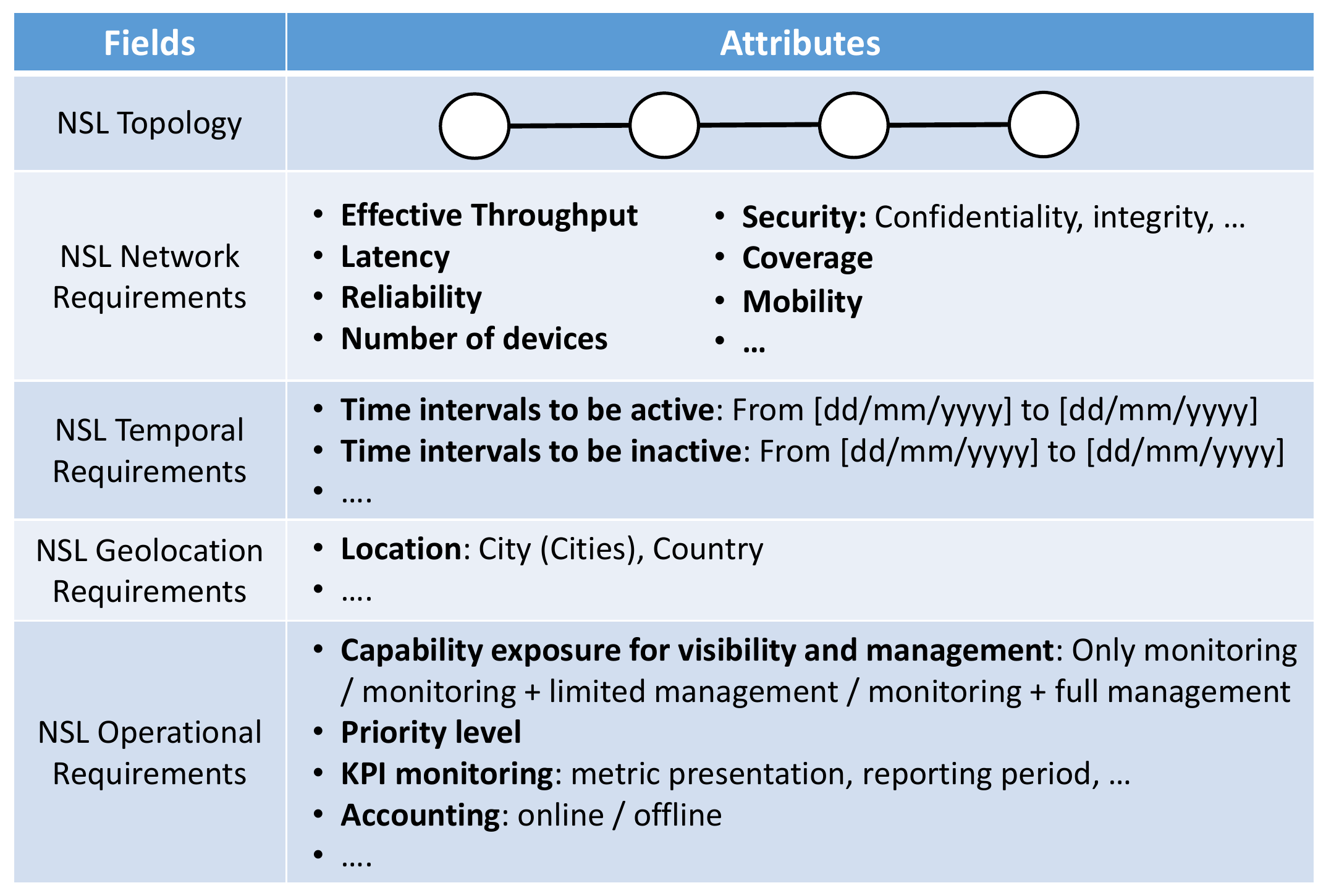}
\caption{Service template structure. The nodes included in the topology depend on the use case the template is designed for (e.g. in an eMBB NSL, some nodes could be a cache, the EPC user plane, and the EPC control plane). Note that the value of some NSL requirements could be specified by the tenant, according to the NSL provider's policies. For typical values in different vertical-driven use cases, please see \cite{NGMN}.}
\label{fig:NSLTemplate}
\end{figure}

The NSL provider defines a business-driven Service Catalog that contains a finite set of service templates, each describing a different service offering. These offerings include NSLs optimized to serve a multitude of usage scenarios, ranging from typical 5G services  (e.g. eMBB, mMTC, and uRLLC) to vertical-specific applications  (e.g. smart factory, remote surgery, connected cars, etc.). A service template is a readymade document that contains all the information that is required to drive the deployment of an NSL. In particular, it contains (1) the NSL topology, expressed as an ordered chain of technology-agnostic composable nodes, each providing specific functionality; (2) the NSL network requirements, including performance and functional requirements; (3) the NSL temporal requirements; (4) the NSL geolocation requirements; and (5) the NSL operational requirements. An example of a service template is shown in Fig. \ref{fig:NSLTemplate}.

To facilitate the customization and automate the service definition, the NSL provider may suggest typical configurations of certain attributes, allowing tenants to focus on the key areas of the service template. The number and diversity of attributes that can be specified (including their allowed value ranges) by the tenant is up to the NSL provider's policies.

To order an NSL, the tenant makes use of the self-ordering APIs that the NSL provider exposes in a self-service Web Portal. With these APIs, the tenant gains access to the Service  Catalog, from which it selects the service template that best matches its needs. Then, the tenant specifies the desired values for the attributes it can customize, according to the NSL provider's policies. The result is a catalog-driven NSL service order that the NSL Orchestrator must process. This order contains information mappable to RAN, transport, and core network domains. For simplicity, we focus on the latter two.

\subsection{Network Slice Resource Description}\label{subsec:nslresourcedescription}

The goal of this stage is to give a resource-centric view of the ordered NSL, expressed through an NSL-IL (see Section \ref{sec:NFV}). This NSL-IL may be used to decide if the ordered NSL is feasible/infeasible from a resource viewpoint, and hence accepted/rejected for deployment (see subsection \ref{subsec:AdmissionControl}). 

Upon receiving the service order, the NSL Orchestrator extracts the content that is relevant from a resource viewpoint: the NSL topology, and the NSL network requirements (i.e. performance and functional requirements). Using this information, the NSL Orchestrator constructs an NSL-IL for the NSL instance. For this end, it performs three steps. 

In the first step, the NSL Orchestrator uses the NSL topology to identify which NS(s) need to be deployed for the NSL, retrieving the corresponding NS descriptor(s) from the NS Catalog. In the second step, the NSL Orchestrator selects within each descriptor the deployment option that best matches the features and the performance level required for the NSL. In other words, it selects the triplet (NS descriptor ID, NS Flavor ID, NS-IL ID) to be used to instantiate each NS. Finally, the NSL Orchestrator constructs the NSL-IL by referencing the selected triplet(s).

\begin{figure}[htbp!]
\centering 
\includegraphics[width=0.45\textwidth]{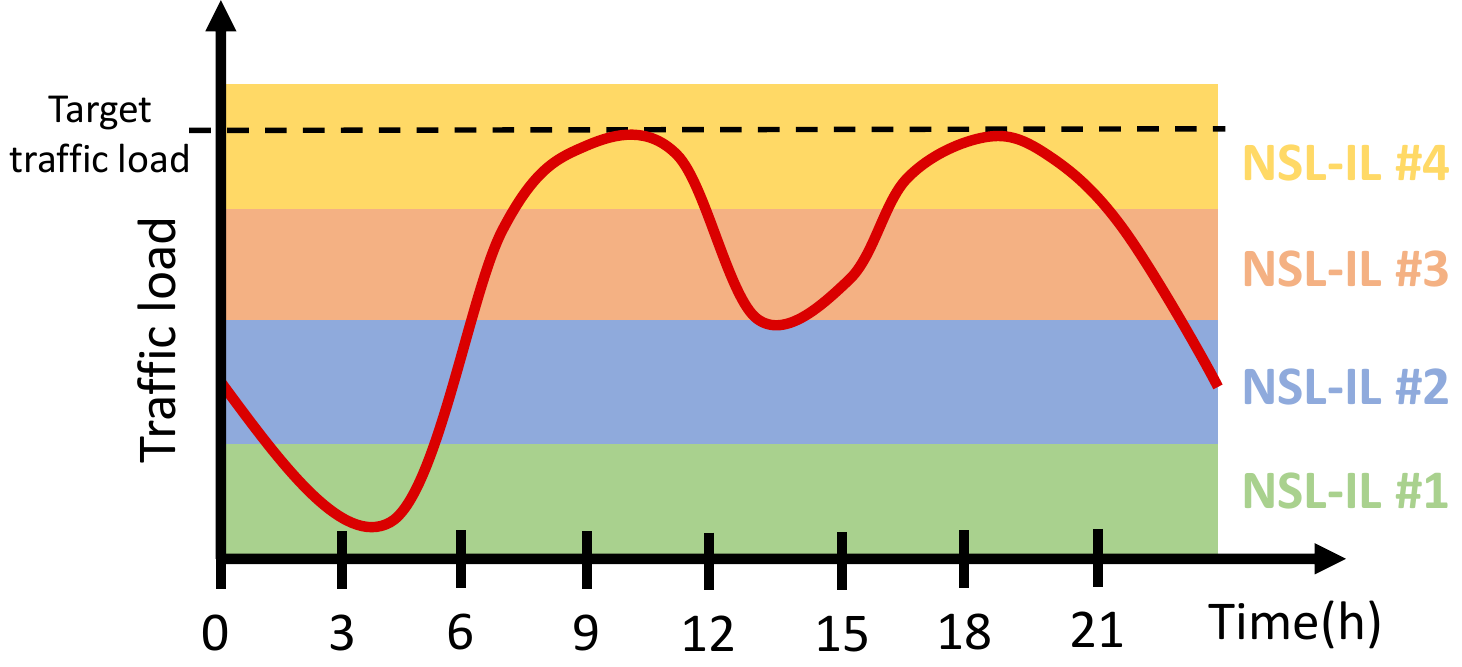}
\caption{Example of the traffic load expected for a given NSL instance  during a typical day. NSL-IL \#4 is the target NSL-IL, and the rest are the optional NSL-ILs. The entire set of NSL-ILs enables the NSL provider to adjust the level of resources within the NSL instance at run-time, in such a way it satisfies the desired performance, while making an efficient resource usage.}
\label{fig:TrafficLoad}
\end{figure}

With the mentioned approach, the constructed NSL-IL meets the specified network requirements of the NSL instance, and hence is able to accommodate the target traffic load. From here on out, we will refer to this NSL-IL as the \textit{target NSL-IL}. However, traffic fluctuations may occur throughout the lifetime of the NSL instance, resulting in periods of time  where the traffic load is considerably lower than the target one. In this kind of situations, the triplet(s) used for the target NSL-IL may lead to a waste of resources. To solve this issue and take advantage of multiplexing gains, the NSL Orchestrator could make use of less resource-demanding triplets to accommodate lower traffic loads, and construct \textit{optional NSL-ILs} with them (see Fig. \ref{fig:TrafficLoad}). The number of optional NSL-ILs and the triplet(s) selected for each of them depend on the traffic fluctuations expected for the NSL instance. To estimate these fluctuations, the NSL Orchestrator may rely on traffic models that the NSL provider has inferred from historical data.

The target NSL-IL, along with the optional NSL-ILs, define the complete set of NSL-ILs among which the NSL instance can scale up/down during its entire life cycle. 

\subsection{Admission Control}\label{subsec:AdmissionControl}
The target NSL-IL specifies the resource requirements fitting the tenant's demands. Once derived, the NSL provider can perform the admission control. The admission control aims to check if the NSL provider can  satisfy the resource, geolocation, and temporal requirements of the ordered NSL. For this end, the following information is needed:

\begin{itemize}
\item [(1)] The resource requirements of the NSL instance. This includes(a) the resources to be allocated for each VNF instance and virtual link, (b) the affinity/anti-affinity rules applicable between VNF instances, and (c) the reliability requirements for each VNF instance and virtual link.
\item [(2)] The geographical region(s) where each VNF is needed.
\item [(3)] The time intervals when the NSL instance needs to be active (operative).
\item [(4)] Information of the PoPs (and the WAN network(s) connecting them) to which the NSL provider is subscribed.
\end{itemize}

The information shown in (1)-(3) is available to the NSL Orchestrator; indeed, (1) is part of the target NSL-IL, while (2)-(3) are derived from the geolocation and temporal requirements specified in the service order. The information specified in (4) is available to the Resource Orchestrator, and provided by the underlying VIM(s)/WIM(s). The fact that the NSL Orchestrator and the Resource Orchestrator operate at different abstraction levels means that they deal with different level of information, and hence none of them is able to perform the admission control at its own. The interplay of both functional blocks is needed. Following this idea, the admission control can be splitted into three steps. The NSL Orchestrator performs the first two steps, being the latter carried out by the Resource Orchestrator.

In the first step, the NSL Orchestrator calculates which PoP(s) is (are) candidate to host each VNF instance. A PoP is candidate for a VNF instance if the location and capabilities of the PoP satisfy the geolocation and reliability requirements of that instance. For this step, the NSL Orchestrator takes as inputs the information specified in (1c) and (2), and the resource-agnostic view provided by the Resource Orchestrator. As seen in Section \ref{sec:run-time}, this view consists of high-level information of the location and capabilities of the reachable PoPs.

In the second step, the NSL Orchestrator sends two kind of data to the Resource Orchestrator. On one hand, data concerning the NSL lifetime. For this end, the NSL Orchestrator takes the information shown in (3), and passes it down to the Resource Orchestrator. On the other hand, data concerning the target NSL-IL to be accommodated. For that, the NSL Orchestrator takes the resource requirements specified in (1), along with the candidate PoPs calculated in the first step, and passes then down to the Resource Orchestrator at VNF/virtual link level. For each VNF instance, the NSL Orchestrator communicates the candidate PoP(s), and the requirements shown in (1a) and (1b). For each virtual link, the NSL Orchestrator communicates the requirements specified in (1a) and (1c).

In the third step, the Resource Orchestrator seeks feasible solutions to deploy the target NSL-IL. A solution is feasible as long as each VNF instance can be allocated in a candidate PoP during the time interval(s) in which the NSL instance needs to be active, while satisfying the VNF affinity/anti-affinity rules and connectivity needs. For this step, the Resource Orchestrator takes the data received  from the NSL Orchestrator, and compares it against the information specified in (4).

 If there exists one feasible solution, the admission control is successful. In this case, the Service Level Agreement between the NSL provider and the tenant can be formalized; otherwise, these two parties shall re-negotiate the content of the service order.

\subsection{Optimization and Resource Reservation}
A successful admission control may derive multiple feasible solutions for the target NSL-IL (e.g. multiple PoPs can accommodate a given VNF instance). However, only one of them must be eventually selected for deployment. To solve this issue, the Resource Orchestrator may run an algorithm that calculates the optimal solution. Examples of optimality criteria that could be used for this algorithm include minimize resource usage, minimize energy consumption, etc.

Once the optimal solution is found, the Resource Orchestrator may proceed with resource reservation. The Resource Orchestrator sends resource reservation requests towards the underlying VIM(s)/WIM(s). The hard and soft nature of this reservation depends on the NSL provider's policies, as well as the nature of the use case the NSL instance will accommodate. 

\subsection{Network Slice Preparation}

The NSL preparation is the last stage prior to put the NSL operative. It consists of setting up all that is required to manage the NSL instance throughout its entire life cycle, from commissioning (instantiation, configuration, and activation) to decommissioning (de-activation and termination) \cite{3gpp:28.801}. This includes (1) preparing the network environment, and (2) designing and on-boarding the NSL descriptor. 

In the network environment preparation, the NSL Orchestrator performs the following tasks:
\begin{itemize}
\item \textbf{It negotiates with the Resource Orchestrator a priority level for the NSL instance}. Having different priority levels allows the Resource Orchestrator to define a priority order between the NSL instances in case they compete for the same resources, or in case of resource scarcity.
\item \textbf{It prepares the management plane of the NSL instance}. First, the NSL Orchestrator instantiates the NSL Manager, the Tenant SDN Controller, the NS Orchestrator, and the VNFM(s). Then, it configures these functional blocks in an appropriate manner, making them ready for the run-time phase. By means of example, the NSL Orchestrator configures the NSL Manager in such a way it provides the tenant only with the visibility and the management capabilities specified in the service order. 
\end{itemize}

In parallel to the network environment preparation, the NSL Orchestrator builds up the NSL descriptor. The NSL descriptor is a deployment template used by the NSL Manager to operate the NSL instance during its life cycle in an agile, automated fashion. This descriptor includes the following parts: 

\begin{itemize}
\item \textbf{A set of policy-based workflows}. These workflows enables the NSL Manager to enforce the expected behavior of the NSL instance during its life cycle, in a timely manner. The NSL Manager translates the content of these workflows into appropriate NS and VNF management actions, and forwards them to the NS Orchestrator and to the tenant SDN controller for their enforcement. 
\item \textbf{The set of NSL-ILs available for use}, constructed in the Network Slice Resource Description phase (see subsection \ref{subsec:nslresourcedescription}). The NS Orchestrator use the triplets referenced by these NSL-ILs to scale the NS instance(s) at run-time, according to time-varying traffic demands. 
\item \textbf{VNF configuration primitives at application level, and VNF chaining management instructions}. Both are used by the Tenant SDN Controller to programmatically configure and chain the VNF instance(s).
\item \textbf{Information about management data}, used for performance management (e.g. metrics to be monitored, metric presentation, reporting period) and fault management (e.g. alarms to be subscribed). Derived from the NSL operational requirements specified in the service order, the management data may be collected from the NS Orchestrator and the Tenant SDN Controller, and used for visibility/manageability purposes.
\end{itemize}

Note that the policy-based workflows contained in the NSL descriptor enables the NSL manager to automate all the life cycle operations that are manually triggered from the OSS in the ETSI NFV framework \cite{nfv:ifa013}, making the NSL instance a self-contained entity. The remaining content of the NSL descriptor is used to feed these workflows (e.g. performance metrics may be taken as inputs for the workflows targeted at the NSL scaling operation).

\section{Conclusions}\label{sec:conclusions}

An SDN/NFV-based network slicing architecture has been presented in this work. This architecture addresses the two phases considered for network slicing: the creation phase and the run-time phase. This work focuses on the former.

We have provided detailed insight into the steps needed to successfully complete catalog-driven NSL deployments. These steps have been arranged into five stages: Service Ordering, Network Slice Resource Description, Admission Control, Optimization \& Resource Reservation, and Network Slice Preparation. In each of these stages, the input/output information required, the steps involved, and the role of the participant functional block(s) have been specified. 

With the architecture and the ordered set of stages proposed in this work, NSL providers are able to perform cost-efficient deployments, in an agile, flexible, and automated manner. The presence of a Service Catalog, with customizable service offerings that brings flexibility in service definition, and the interplay between the NSL Orchestrator and Resource Orchestrator are crucial for that end. Additionally, the correct design of a NSL descriptor in the creation phase is key for a successful operation in the run-time phase. This descriptor makes the NSL instance a self-contained entity, enabling the slice-specific management plane to operate the NSL instance in a customized way, with great agility, and full automation.


%


\section*{Acknowledgment}
This work is partially supported by the Spanish Ministry of Economy and Competitiveness and the European Regional Development Fund (Project TEC2016-76795-C6-4-R), the Spanish Ministry of Education, Culture and Sport (FPU Grant 16/03354), and the University of Granada, Andalusian Regional Government and European Social Fund under Youth Employment Program.

\ifCLASSOPTIONcaptionsoff
  \newpage
\fi



%

\bibliographystyle{ieeetr}
\bibliography{main}


  


%





\end{document}